\documentclass{amsart}
\usepackage{graphicx}
\theoremstyle{definition}

\theoremstyle{remark}

\numberwithin{equation}{section}

\newcommand{\To}{\longrightarrow}

\begin{document}
\title{Hopf Algebra Structure of a Model Quantum Field Theory}%
\author{A. I. Solomon$^{a,b}$,G. E. H. Duchamp$^{c}$, P. Bl{}asiak$^{d}$,A. Horzela$^{d}$,
K. A. Penson$^{b}$}%
\address{
$^{a}$The Open University, Physics and Astronomy Department\\
Milton Keynes MK7 6AA, United Kingdom\\
$^{b}$Lab.de Phys.Th\'eor,de la Mati\`ere Condens\'ee, CNRS UMR 7600,
Universit\'e Pierre et Marie Curie\\
Tour 24 -- 2e \'et., 4 Pl.Jussieu, F 75252 Paris Cedex 05, France\\
$^{d}$H.Niewodnicza\'nski Institute of Nuclear Physics, Polish Academy of Sciences\\
ul. Eliasza-Radzikowskiego 152, PL 31342 Krakow, Poland\\
$^{c}$ Institut Galil´ee, LIPN, CNRS UMR 7030 99 Av. J.-B. Clement, F-93430 Villetaneuse,
France}%
\email{ a.i.solomon@open.ac.uk, blasiak@lptl.jussieu.fr, gduchamp2@free.fr,
 andrzej.horzela@ifj.edu.pl, penson@lptl.jussieu.fr}%


\dedicatory{Talk presented at G26 International Conference on Group Theoretical
 Methods in Physics, New York, June 2006.}%
\begin{abstract}
Recent elegant work\cite{kre} on the structure of Perturbative Quantum Field Theory (PQFT) has
revealed an astonishing interplay between analysis(Riemann Zeta functions), topology
(Knot theory), combinatorial graph theory (Feynman Diagrams) and algebra (Hopf
structure). The difficulty inherent in the complexities of a fully-fledged field theory
such as PQFT means that the essential beauty of the relationships between these areas can
be somewhat obscured. Our intention is to display some, although not all, of these
structures in the context of a simple zero-dimensional field theory; i.e. a quantum theory
of non-commuting  operators which do not depend on spacetime. The combinatorial properties of these
boson creation and annihilation operators, which is our chosen example, may be described
by graphs \cite{bbm,bbs}, analogous to the Feynman diagrams of PQFT, which we show possess a
Hopf
algebra structure\cite{geh}. Our approach is based on 
the partition function for a  boson gas. In a subsequent note in these Proceedings we 
sketch the relationship between the Hopf algebra of our simple model and that of the PQFT
algebra.
\end{abstract}
\maketitle
\section{Partition Function Integrand}
Consider the Partition Function $Z$ of a quantum statistical mechanical system 
\begin{equation}\label{pf1}
Z={\rm Tr}\exp(-\beta H)\,.
\end{equation}
whose  hamiltonian is $H$  ($\beta \equiv 1/kT$, $k$=Boltzmann's constant $T$=absolute temperature).
We may evaluate the trace over any complete set of states; we choose the (over-)complete set of coherent states
\begin{equation}\label{cs}
|z\rangle= e^{-|z|^2|/2}\sum_n ({{z^n}{/n!}) {a^{\dagger}}^n}|0\rangle 
\end{equation}
where $a^{\dagger}$ is the boson creation operator satisfying $[a.a^{\dagger}]=1$,
for which the completeness or {\em resolution of unity}
property is
\begin{equation}\label{cs1}
 \frac{1}{\pi}\int d^{2}z |z\rangle\langle z|=I\equiv
 \int d\mu(z)|z\rangle\langle z|.
 \end{equation}
 The simplest, and generic, example is the free single-boson hamiltonian 
 $H=\epsilon a^{\dagger}a$ 
 for which the appropriate trace calculation is
\begin{eqnarray}\label{tr1}
Z&=&\frac{1}{\pi}\int d^{2}z \langle z|\exp\bigl(-\beta \epsilon a^{\dagger}a\bigr)|z\rangle=\nonumber \\
&=&\frac{1}{\pi}\int d^{2}z \langle
z|:\exp\bigl(a^{\dagger}a(e^{-\beta\epsilon}-1)\bigr):|z\rangle.
\end{eqnarray}
Here we have used the  well-known relation
\cite{[5],[6]} for the {\em forgetful} normal ordering operator
$:\!f(a, a^{\dagger})\!:$ which means ``normally order the
creation and annihilation operators in $f$ {\em forgetting} the
commutation relation $[a,a^{\dagger}]=1.$''\footnote{Of course,
this procedure may alter the value of the operator to which it is
applied.}

We may write the Partition Function in general as 
\begin{equation}\label{pf2}
Z=\int{F(x,z)\,d\mu(z)}\\; \; \; \; \; (x\equiv -\beta \epsilon)
\end{equation}
thereby defining the Partition Function Integrand (PFI) $F(x,z)$, which will be the object of our analysis.
\section{Combinatorial aspects: Bell numbers}
The  generic free-boson example Eq.(\ref{tr1}) above may be rewritten  to show the connection with certain well-known combinatorial numbers. Writing $y=|z|^2$, Eq.(\ref{tr1}) becomes
\begin{equation}\label{z2}
Z=\int\limits_0^\infty dy\exp\bigl(y(e^{x}-1)\bigr)\,.
\end{equation}
This is an integral over  the classical {\em exponential generating function} for
the {\em Bell polynomials}
\begin{equation}\label{bpgf1}
 \exp\left(y\bigl(e^{x}-1\bigr)\right)=\sum_{n=0}^\infty B_n(y)\,\frac{x^n}{n!}\
\end{equation}
where the Bell number is $B_{n}(1)=B(n)$, the number of ways of putting $n$
different objects into $n$ identical containers (some may be left
empty). Related to the Bell numbers
are the {\em Stirling numbers of the second kind} $S(n,k)$, which
are defined as the number of ways of putting $n$ different objects
into $k$ identical containers, leaving none empty. From the
definition we have
$B(n)=\sum_{k=1}^n S(n,k)$.
The foregoing gives a 
 combinatorial interpretation of  the partition function integrand $F(x,y)$ as 
the exponential generating function of
the  Bell polynomials.
\subsection{Graphs}
We now give a graphical representation of the Bell numbers.
Consider labelled lines which emanate from a white dot, the
origin, and finish  on a black dot, the vertex. We shall allow
only one line from each white dot but impose no limit on the
number of lines ending on a black dot. Clearly this simulates the
definition of $S(n,k)$ and $B(n)$, with the white dots playing the
role of the distinguishable objects, whence the lines are
labelled, and the black dots that of the indistinguishable
containers. The identification of the graphs for 1, 2 and 3 lines
is given  in Figure 1.
\begin{figure}[h]
\hspace{1cm}
\resizebox{10cm}{!}{\includegraphics{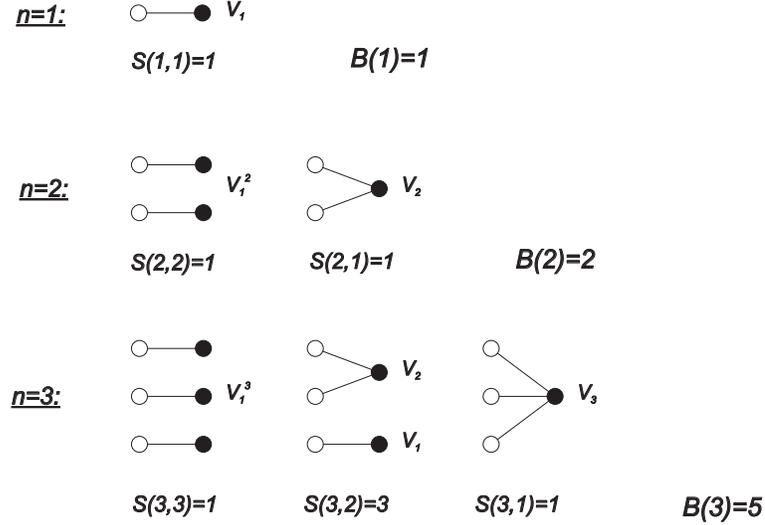}}
\caption{\label{inter}Graphs for $B(n)$, $n=1$, 2, 3.}
\end{figure}

We have concentrated on the Bell number sequence and its
associated graphs since, as we shall show, there is a sense in
which this sequence of graphs is {\em generic}. That is, we can
represent {\em any} combinatorial sequence by the same sequence of
graphs as in  Figure 1, with suitable vertex multipliers
(denoted by the $V$ terms in the same figure). Consider the general partition function
of Eq.(\ref{pf1})
with hamiltonian given by  $H=\epsilon
w(a,a^{\dagger})$, with $w$
 a string (= sum of products of positive powers) of boson
creation and annihilation operators. The partition function
integrand $F$ of Eq.(\ref{pf2}) for which we seek to give a graphical expansion, is now 
\begin{eqnarray}\label{pf3}
F(x,z)&=&\langle z|\exp(xw)|z\rangle= \hskip20mm(x=-\beta \epsilon)\nonumber \\
&=&\sum_{n=0}^{\infty}\langle z|w^n|z\rangle\,\frac{x^n}{n!}=\nonumber \\
&=&\sum_{n=0}^{\infty}W_n(z)\,\frac{x^n}{n!}=\nonumber \\
&=&\exp\biggl(\;\sum_{n=1}^{\infty}V_n(z)\,\frac{x^n}{n!}\biggr),
\end{eqnarray}
with obvious definitions of $W_n$ and $V_n$. The sequences
$\left\{W_n\right\}$ and $\left\{V_n\right\}$ may each be
recursively obtained from the other \cite{[7a]}. This relates the
sequence of multipliers $\{V_n\}$ of Figure 1 to the hamiltonian
of Eq.(\ref{pf1}). The lower limit $1$ in the $V_n$ summation is a
consequence of the normalization of the coherent state
$|z\rangle$.
A mild generalization is to write the Partition Function Integrand of Eq.(\ref{pf3}), using the product formula \cite{bbs,Vasiliev, ben1}, as 
\begin{equation}\label{pf4}
F(x,{\mathbb{V}},{\mathbb{L}})=\left.\exp\left(\sum_{m=1}^{\infty} L_m \frac{x^m}{m!}\frac{d^m}{dy^m}\right)
 \exp\left(\sum_{s=1}^{\infty} V_s \frac{y^s}{s!}\right)\right|_{y=0} .
\end{equation}
(We have suppressed the explicit dependence on the coherent state parameter $z$.) The advantage of this formulation is that it treats the white and black spots symmetrically\footnote{Thus we need not adhere to the previous convention of treating the white spots as the origins.}, as well as having some calculational advantages.  An example of some of the associated graphs is given in Figure 2.
\section{Hopf Algebra structure}
We briefly describe the  Hopf algebra   ${\bf BELL}$ which  the diagrams of Figure 1 define.  
\begin{enumerate}
\item Each distinct diagram is an individual basis element of ${\bf BELL}$; thus the dimension is infinite. (Visualise each diagram in a ``box''.) The sum of two diagrams is simply  the two boxes containing the diagrams. Scalar multiples are formal; for example, they may be  provided by the $V$ coefficients.
\item The identity element $e$ is the empty diagram (an empty box). 
\item Multiplication is the juxtaposition of two diagrams within the same ``box''. ${\bf BELL}$ is generated by the {\em connected} diagrams; this is a consequence of the Connected Graph Theorem \cite{FU}.  Since we have not here specified an order for the juxtaposition, multiplication is commutative. 
\item The coproduct $\Delta:{\bf BELL}\To {\bf BELL}\times {\bf BELL}$ is defined by 
\begin{eqnarray}
\Delta(e)&=&e\times e  \; \; \; \;({\rm unit}\; \; e) \nonumber \\
\Delta(x)&=&x \times e +e \times x  \; \; \; \; ({\rm generator}\; \; x) \nonumber \\
\Delta(AB)&=&\Delta(A)\Delta(B) \; \; \; {\rm otherwise} \nonumber 
\end{eqnarray}
so that $\Delta$ is an algebra homomorphism.
\item The co-unit $\epsilon$ satisfies $\epsilon(e)=1$ otherwise $\epsilon(A)=0$.
\item The antipode ${\mathcal S}:{\bf BELL}\To {\bf BELL}$ satisfies ${\mathcal S}(e)=e$; on a generator $x$, ${\mathcal S}(x)=-x$. It is an {\em anti-homomorphism}, i.e. ${\mathcal S}(AB)={\mathcal S}(B){\mathcal S}(A)$.
\end{enumerate}
It may be shown that the foregoing structure ${\bf BELL}$ satisfies the axioms of a commutative, co-commutative Hopf algebra.
Diagrams such as those of Figure 2, associated with the formulation Eq.(\ref{pf4}) similarly give rise to a commutative, co-commutative Hopf algebra ${\bf DIAG}$ generated by the connected graphs. The Bell Hopf algebra ${\bf BELL}$ is a homomorphic image of ${\bf DIAG}$.
\begin{figure}[h]
\hspace{1cm}
\resizebox{10cm}{!}{\includegraphics{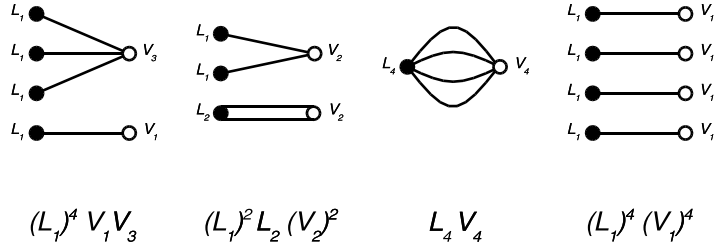}}
\caption{\label{inter2}Some examples of 4-line graphs.}
\end{figure}
\section{Discussion}
The philosophy of our approach has been the following.  To elucidate the structure of a  complicated physical system, such as perturbative quantum field theory (PQFT) , is itself a daunting task.  One may gain some insight by considering a much more straightforward system, such as the one treated  here.  This may be thought of as a zero-dimensional field theory, as our simple boson operator $a$ does not depend on space or time. Nevertheless, we have shown that even such a   basic system does exhibit some of the features of the more complicated case, in particular the structure of the  Hopf algebras 
${\bf BELL}$ and ${\bf DIAG}$. This may be thought of as a simple solvable model in its own right.  However, one may also adopt the approach of asking  wherein does this simple structure sit within the full PQFT structure?  Our approach is to generalize the {\em algebraic} structure, and thereby produce Hopf algebras of sufficient complexity to emulate those associated with PQFT.  

In the following note \cite{GHE} we show how such a generalization may be achieved, with starting point the Hopf algebras ${\bf BELL}$ and ${\bf DIAG}$ described here.


\begin{thebibliography}{999}
\bibitem{kre} A readable account may be found in Dirk Kreimer's
"Knots and Feynman Diagrams", Cambridge Lecture Notes in Physics, CUP (2000).
\bibitem{bbm} We use the graphical description of
Bender, Brody and Meister, J.Math.Phys. {\bf 40}, 3239 (1999) and arXiv:quant-ph/0604164
\bibitem{bbs} This approach is extended in
Blasiak, Penson, Solomon, Horzela and Duchamp, J.Math.Phys. {\bf 46}, 052110 (2005) and
arXiv:quant-ph/0405103
\bibitem{geh} A preliminary account of the more mathematical aspects of this work may be found
in arXiv: cs.SC/0510041
\bibitem{[5]}
J.R. Klauder and E.C.G. Sudarshan: \textit{Fundamentals of Quantum
Optics}. Benjamin, New York, 1968.
\bibitem{[6]}
W.H. Louisell: \textit{Quantum Statistical Properties of
Radiation}. J. Wiley, New York, 1990.
\bibitem{[7a]}
M. Pourahmadi: Amer. Math. Monthly \textbf{91}, 303, 1984.
\bibitem{Vasiliev} N.A. Vasiliev,  
\emph{ Functional Methods in
Quantum Field Theory and Statistical Physics}, Gordon and Breach Publishers, Amsterdam, 1998.
\bibitem{ben1} C.M. Bender,D.C. Brody and B.K.Meister, 
Quantum field theory of partitions, {\it J.Math. Phys.} {\bf 40} 3239,1999;
 C.M. Bender, D.C. Brody and B.K. Meister, 
 Combinatorics and field theory, {\it Twistor Newsletter} {\bf 45} 36, 2000.
\bibitem{FU} G.W. Ford and G.E. Uhlenbeck, Proc. Nat. Acad. {\bf 42}, 122,1956.
 \bibitem{GHE} G.H.E. Duchamp {\em et al}, "A multipurpose Hopf deformation of the Algebra of Feynman-like Diagrams", in these Proceedings.
\end{thebibliography}
\end{document}